# All-optical trion generation in single walled carbon nanotubes

Silvia Santos[1], Bertrand Yuma[2], Stéphane Berciaud[2], Jonah Shaver[1], Mathieu Gallart[2], Pierre Gilliot[2], Laurent Cognet[1] and Brahim Lounis[1*]

[1]*Laboratoire Photonique Numérique et Nanosciences, Université de Bordeaux, Institut d'Optique Graduate School & CNRS, 351 cours de la libération, 33405 Talence, France*
[2]*IPCMS, UMR 7504, CNRS - Université de Strasbourg,*
*23, rue du Lœss, 67034 Strasbourg, France*

* blounis@u-bordeaux1.fr

We present evidence of all optical trion generation and emission in undoped single walled carbon nanotubes (SWCNTs). Luminescence spectra, recorded on individual SWCNTs over a large CW excitation intensity range, show trion emission peaks red-shifted with respect to the bright exciton peak. Clear chirality dependence is observed for 22 separate SWCNT species, allowing for determination of electron-hole exchange interaction and trion binding energy contributions. Luminescence data together with ultrafast pump probe experiments on chirality sorted bulk samples suggest that exciton-exciton annihilation processes generate dissociated carriers that allow for trion creation upon a subsequent photon absorption event.

It is well established that the optical properties of semiconducting single walled carbon nanotubes (SWCNTs) are dominated by tightly bound one-dimensional excitons[1, 2]. The recombination of these excitons results in narrow emission peaks. As it dominates the luminescence spectra, the lowest lying optically active exciton in SWCNTs has attracted much experimental and theoretical attention over the past decade[3]. Since SWCNTs consist only of surface atoms, exciton dynamics, and hence nanotube luminescence intensity, are extremely sensitive to the local environment and the presence of quenching sites or structural defects[4, 5]. Such extrinsic effects are thought to be responsible for the fairly short luminescence lifetimes (100 ps or less[4, 6]) and low luminescence quantum yields (of a few percent) reported for individual SWCNTs in the low excitation intensity regime[6, 7].

For sufficiently strong *CW* or pulsed excitation intensities multiple excitons are likely to exist simultaneously. In this regime, nanotube photophysics is governed by the interplay between one-dimensional exciton diffusion along the nanotube sidewall[5] and strong coulomb interactions between carriers. When two or more excitons are present, their interaction can lead to two different situations. On one hand, they can undergo Auger processes such as exciton-exciton annihilation (EEA) which open-up additional, efficient nonradiative recombination pathways. Experimental evidence for this has been reported by several groups using ultrafast spectroscopy on ensemble samples[8, 9 ]. On the other hand, excitons may form many-body bound states. Theorists have predicted biexcitons to have a binding energy larger than the thermal energy $k_BT$ [10, 11]. No experimental evidence for these bound states has been reported so far, however, in agreement with recent theoretical studies[12].

Charged excitons, or trions, are another class of many-body bound states predicted to possess a significant binding energy in SWCNTs[13]. Very recently, Matsunaga *et al.*[14] observed new spectral features in p-doped nanotube suspensions, which have been assigned to hole-exciton bound states (positively charged trions).

Here we show that trions can be efficiently generated, on demand and in-situ, in highly luminescent *undoped* carbon nanotubes through control of photo-excitation intensity. Trion emission below the main exciton peaks of individual SWCNTs, belonging to 22 different chiralities, is observed. In addition, a transient induced absorption feature is recorded at the trion spectral position in pump-probe spectroscopy experiments on a chirality-sorted (6,5) SWCNT suspension. Both single nanotube luminescence and transient absorption data support a scenario in which localized trions are formed.

Highly luminescent HiPco SWCNTs were used for our single molecule studies. The SWCNTs were first dispersed in aqueous surfactant sodium deoxycholate (DOC) solutions and then either immobilized in agarose gels (5% wt) or spin-coated on surfaces pre-coated with PVP (polyvinylpyrrolidone). Control experiments on SWCNTs in different surfactants (cetyltrimethylammonium bromide, sodium dodecylbenzene sulfonate, or pluoronic) were performed with similar results. For pump-probe experiments, we used a solution of nonlinear density gradient ultracentrifugation (DGU) sorted (6,5) nanotubes suspended in sodium cholate[15].

Wide-field and confocal photoluminescence microscopes were used to image individual SWCNTs belonging to many different (n,m) chiralities. SWCNTs were optically excited with tunable cw lasers tuned near their second order resonance $S_{22}$ or at their K-momentum exciton-phonon sideband[16] [17]. Individual SWCNTs were imaged with a InGaAs camera and their luminescence spectra were recorded with a spectrometer equipped with a InGaAs detector array. These nanotubes showed bright, highly stable emission with the longest luminescence decays reported to date, signifying low defect density and reduced environmental perturbations[18].

Bright individual nanotubes were selected (see inset of figure 1a) and their luminescence spectra were first recorded at excitation intensities in the monoexcitonic regime (well below 1 kW/cm²) (figure 1a-c). In addition to the bright singlet-exciton peak (X), the phonon sideband of the K-momentum dark exciton is clearly visible at a nearly chirality independent shift of ~130 meV [17, 19].

At higher excitation intensities ( > kW/cm²) (Fig. 1 d-f), a new emission peak is systematically observed, thereafter denoted X*. The X* peak is red-shifted with respect to the X peak. However, unlike the phonon sideband, the energy shift ∆E between bright exciton and X* exhibits a strong chirality dependence. The X* peaks are distinctively red-shifted by at least 50 meV as compared to those observed in defect-induced brightening of dark triplet excitons [19-21] and their energies differ significantly from the values observed in oxygen doped SWCNTs [22]. Furthermore, X* emission appears only at high laser intensities and is absent when returning to the low excitation regime, whereas emission from brightened triplet excitons or oxygen doped SWCNTs is observed at any laser intensity following strong pulsed-laser irradiation [19-21] or exposure to ozone [22].

Our setup allowed us to determine ∆E's dependence on SWCNT diameter and chiral angle with excellent accuracy (figure 2). Each point of figure 2 represents the peak value of the ∆E distribution obtained for a given chirality (as exemplified in the inset of figure 2). The distributions contained more than hundred individual SWCNTs for the most abundant chiralities and at least five for minority species. $\Delta E$ tends to increase with decreasing SWCNT diameter and displays a clear family pattern, reminiscent of the well-known Kataura plot. For a given SWCNT family, with a fixed value of (2n+m), deviations from a main trend increase as chiral angle decreases (armchair to zigzag). As observed for bright singlet exciton binding energies, these deviations are much more pronounced for the [mod(n-m,3)=1] families. This family pattern strongly suggests that the X* peak stems from an intrinsic property of SWCNTs, and, as they are at the same energies as in reference [14], we assign the X* peaks to trions. In contrast to Matsunaga *et al.*, however, no chemical doping is involved in our study, the carriers being photo-generated.

Trion binding energy in SWCNTs is predicted to scale as 1/d, (d being the tube diameter) and to range, for (6,5) nanotubes, between 50 meV and 132 meV depending on dielectric screening, $\varepsilon$=4 and $\varepsilon$=2, respectively [13]. However, $\Delta E$~190 meV measured here for (6,5) nanotubes is found to be significantly larger than the predicted range of $\Delta E$ assuming reasonable values of $\varepsilon$ for surfactant-wrapped SWCNTs in aqueous environments. Moreover, the diameter-dependence of $\Delta E$ for near armchair SWCNTs cannot be fitted with an A/d function, as would be expected for a trion formed from a hole and a singlet exciton[13]. Assuming that $\Delta E$ includes a contribution from exchange interaction, an additional term is needed. A $1/d^2$ scaling has been predicted for the singlet-triplet splitting that arises due to the exchange interaction [23]. Indeed we found that a $A/d+B/d^2$ functional form captures the diameter dependence of $\Delta E$ (see Fig. 2), suggesting $\Delta E$ contains a contribution from the singlet-triplet exciton splitting in addition to the trion binding energy. The fitted trion binding energy constant, *A*=85 meV.nm, is consistent with a dielectric constant of 2.2 according to reference [13], close to that experienced by nanotubes in our experiment. The fitted exchange interaction constant, *B*=48 meV.nm², is in agreement with the theoretical value [23]. Furthermore, as expected, this *B* value for surfactant wrapped nanotubes is lower than previously deduced from triplet brightening experiments performed on suspended nanotubes, the latter experiencing lower dielectric screening [14, 21].

In order to investigate the mechanism leading to trion emission, we studied the laser intensity dependence of the X and X* peaks in individual (6,5) SWCNTs (figure3). Long, uniformly bright nanotubes were selected (see figure 1a) and distinct segments (~0.4-0.5 µm long) of the same nanotubes, a few micrometers away from one another, were used to perform intensity dependent studies at different photon enrgies. SWCNTs were excited either in the visible at 2.21 eV, near the $S_{22}$ resonance or in near-IR at 1.47 eV, corresponding to the exciton phonon sideband. In order to have a quantitative comparison between the two different excitation schemes, we carefully evaluated the number of absorbed photons, $N_{ex}$, (or prepared excitons) per unit time and unit length in the nanotubes in both cases using published photophysical parameters [6, 15]. In the spectra, X* emission emerges from background at $N_{ex} \sim 100$ excitons $\mu m^{-1}.ns^{-1}$, a value which corresponds to a few (1 to 2) excitons present on average in a diffusion length (~100-200 nm, after taking account an exciton lifetime of ~ 100ps [5, 18]). This strongly indicates that the third carrier constituting the trion comes from a process of exciton-exciton interaction resulting in the annihilation of one exciton and the dissociation of the second one (see proposed mechanism on fig 3e).

One key question is whether the photo-created charge carriers are freely diffusing along the SWCNT sidewalls or if they are localized. In the latter scenario, are these carriers trapped permanently or transiently present in the nanotubes?

At low $N_{ex}$ (<10 excitons $ns^{-1}.\mu m^{-1}$) X emission exhibits a similar linear rise for the two excitation wavelengths, at higher $N_{ex}$, however, they differ drastically. When excited near the $S_{22}$ resonance, X emission is reproducible as long as the excitation remains sufficiently low. It reaches its maximum at $N_{ex}$~10's of excitons $\mu m^{-1}.ns^{-1}$ before saturating and even decreasing irreversibly for higher exciton densities. When exciting in the near-IR, the X emission is reversible across the whole $N_{ex}$ range studied here and reaches much higher rates before leveling off at greater $N_{ex}$ (several hundreds of excitons $ns^{-1}.\mu m^{-1}$) (figure 3a). In contrast to the X peak, the X* emission rate keeps increasing for $N_{ex}$ up to ~1500 excitons $\mu m^{-1}ns^{-1}$, and exhibits a similar evolution with $N_{ex}$, regardless of the excitation photon energy (figure 3b). The saturation of X emission can have two different origins: EEA, which occurs on the ~1 ps timescale for two excitons prepared in ~0.4 µm nanotube segments[8, 24], or diffusion limited contact-induced nonradiative relaxation of excitons at stationary quenching sites [5]. The former process is intrinsic and does not depend on the excitation wavelength [25] since the prepared excitons relax extremely rapidly down to the emitting state before interacting[26]. Thus, EEA cannot explain the observed difference in the evolutions of the X emission rates with increasing $N_{ex}$ of figure 3a. The latter, however could depend on the excitation photon energy since the luminescence quenching sites can be laser-induced[27]. Indeed, in comparison to near-IR excitation, $S_{22}$ excitation presumably populates more photo-reactive excitonic levels and thus produces a greater number of quenching sites[28] which dramatically degrade X emission (figure 3a). Accordingly, when returning to lower intensity after high illumination, the initial X count rate is not recovered (not shown) with $S_{22}$ excitation in contrast to near-IR excitation (black symbol in figure3a-b). This indicates that the saturation behavior for near-IR excitation is mainly due to EEA while for $S_{22}$ excitation it is dominated by laser induced quenching sites.

The fact that, unlike X emission, X* emission shows similar dependence on $N_{ex}$, regardless of the excitation wavelength, suggests that trion localization prevents their migration to photo-induced quenching sites. We conjecture that localization arises from electrostatic potential fluctuations[29] induced by inhomogeneities in the nanotube environment[30, 31]. Trion localization is further supported by the Lorentzian shape of the X* emission line (Fig. 3d). Indeed, the recombination of a free trion would produce a free carrier and give rise to an asymmetric emission line with a tail towards the low photon energies due to the effective mass difference between the initial and final quasi-particles[32].

In order to investigate the dynamics of these localized charge carriers that facilitate trion formation, we performed time-resolved pump-probe measurements on sorted, single chirality (6,5) nanotube suspensions[15]. Femtosecond, near-IR laser pulses, delivered by a regenerative Ti:Sapphire amplifier with a repetition rate of 200 kHz were used to pump an optical parametric amplifier (OPA) and a sapphire crystal. The pump pulses, with a photon energy of 1.27 eV, in resonance with the $S_{11}$ transition, and fluences in the range of $5.10^{12} - 2.5\ 10^{14}$ photons per $cm^2$ were delivered by the OPA. Continuum probe pulses covering the X* spectral range were generated in the sapphire crystal. Differential transmission spectra were acquired on a spectrometer-coupled InGaAs linear array as a function of the time-delay between the pump and probe pulses, with a temporal resolution ~100 fs, and a sensitivity down to a few $10^{-5}$.

In figure 4a, a clear induced absorption (IA) feature is visible at the X* position of (6,5) SWCNTs. The resonant pump pulses are sufficiently intense to generate multiple excitons in the SWCNTs. These excitons undergo fast EEA, resulting in the creation of a population of charge carriers that provide the third body for trion formation following absorption of probe photons at 1.07 eV (see fig 3e). The X* induced absorption signal starts rising instantaneously after pump excitation and reaches its maximum within a few picoseconds, as shown on Fig. 4b. This X* rise time is comparable to the first short decay time of the X population [33] due to EEA processes. The decay time of the X* induced absorption reflects the evanescence of the photo-generated charge carrier population within several hundreds of picoseconds. For trion formation to occur, photo-induced charge carriers must spatially separate and should thus be localized to avoid spatial overlap, which would lead to binding and further recombination. The dynamics observed in transient absorption experiments (see fig 4b) would then reflect the creation and trapping of the charge carriers (IA rise), and their untrapping (IA decay). The fairly slow relaxation displayed in figure 4 further supports a scenario in which the charge carriers are indeed localized. Importantly, the existence of a transient signal at positive delays and its absence at negative delays demonstrates that the carriers that allow trion formation are indeed photo-generated and not present permanently in the SWCNTs.

In conclusion, we have shown trion emission in pristine carbon nanotubes. We predict that investigations of trion photophysics in various media and temperature conditions could be used as a probe of electrostatic environment, a key parameter for the use of carbon nanotubes in photonic and opto-electronic devices. This work also opens the possibility of all optical manipulation of electron or hole spins in SWCNTs for applications in quantum information[34].

Acknowledgements

We warmly thank R. B. Weisman and S. Ghosh for supplying us with the DGU samples and critical reading of the manuscript. We also thank Jean Besbas for experimental help and Bernd Hönerlage for helpful discussions. This work was funded by ANR, Région Aquitaine, and the European Research Council. JS acknowledges support from the William J. Fulbright Commission.

**Figure captions:**

Figure 1: (color online) Luminescence spectra of bright individual SWCNTs recorded at low CW excitation intensities (mono-excitonic regime) of different chiralities (a-c). Regions of the spectra are expended in order to highlight the K-momentum exciton phonon sidebands which are systematically found at ~135meV from the X line. Inset of (a) shows the luminescence image of a typical uniformly bright nanotube used in this study. (d-f): luminescence spectra of the same nanotubes as in (a-c) recorded under high CW excitation intensities. A clear chirality dependent emission line, denoted X*, appears redshifted from the X line.

Figure 2: (color online) Chirality dependence of the energy difference $\Delta E$ between the X and X* peaks. Nanotubes belonging to the same (2n+m) families are displayed with the same colors as indicated on the figure. The dashed curve is a fit of the near armchair nanotube's $\Delta E$ with $A/d+B/d^2$ yielding A=85 meV.nm and B=48 meV.nm². The top inset shows the two contributions to $\Delta E$, i.e. the exchange interaction energy $B/d^2$ and the trion binding energy $A/d$. Histogram presented in the inset represents the $\Delta E$ values obtained from 125 (6,5) nanotubes.

Figure 3: (color online) (a) Dependence of the bright exciton (X) emission rate recorded on an individual (6,5) nanotube as a function of $N_{ex}$ (excitons prepared per unit length and time) for increasing excitation and for two different excitation wavelengths. The nanotube was excited near the $S_{22}$ transition (green) and at the K-momentum exciton phonon sideband (red) . (b) Zoom of (a) at low $N_{ex}$. (c) $N_{ex}$ dependence of the trion (X*) emission rate. (d) Emission spectrum (black) of the (6,5) nanotube recorded using $N_{ex}$~500ns$^{-1}$.µm$^{-1}$ ($S_{22}$ excitation) fit by a Lorentzian profile (green). (e) Proposed mechanism for trion formation. Under high excitation rates, exciton-exciton annihilation creates a population of charge carriers transiently trapped at local heterogeneities of the electrostatic potential induced by the nanotube environment. Subsequent photon absorption leads to trion formation upon binding of one exciton to one of these localized carriers.

Figure 4: (color online) Transient absorption spectra measured with DGU sorted (6,5) nanotubes at 4 different pump-probe delays. The pump beam was resonant with the bright exciton (X) (1.26 eV). A clear induced absorption is observed at the X* line position. (b) Transient dynamics at the trion (X*) induced absorption (1.065eV).

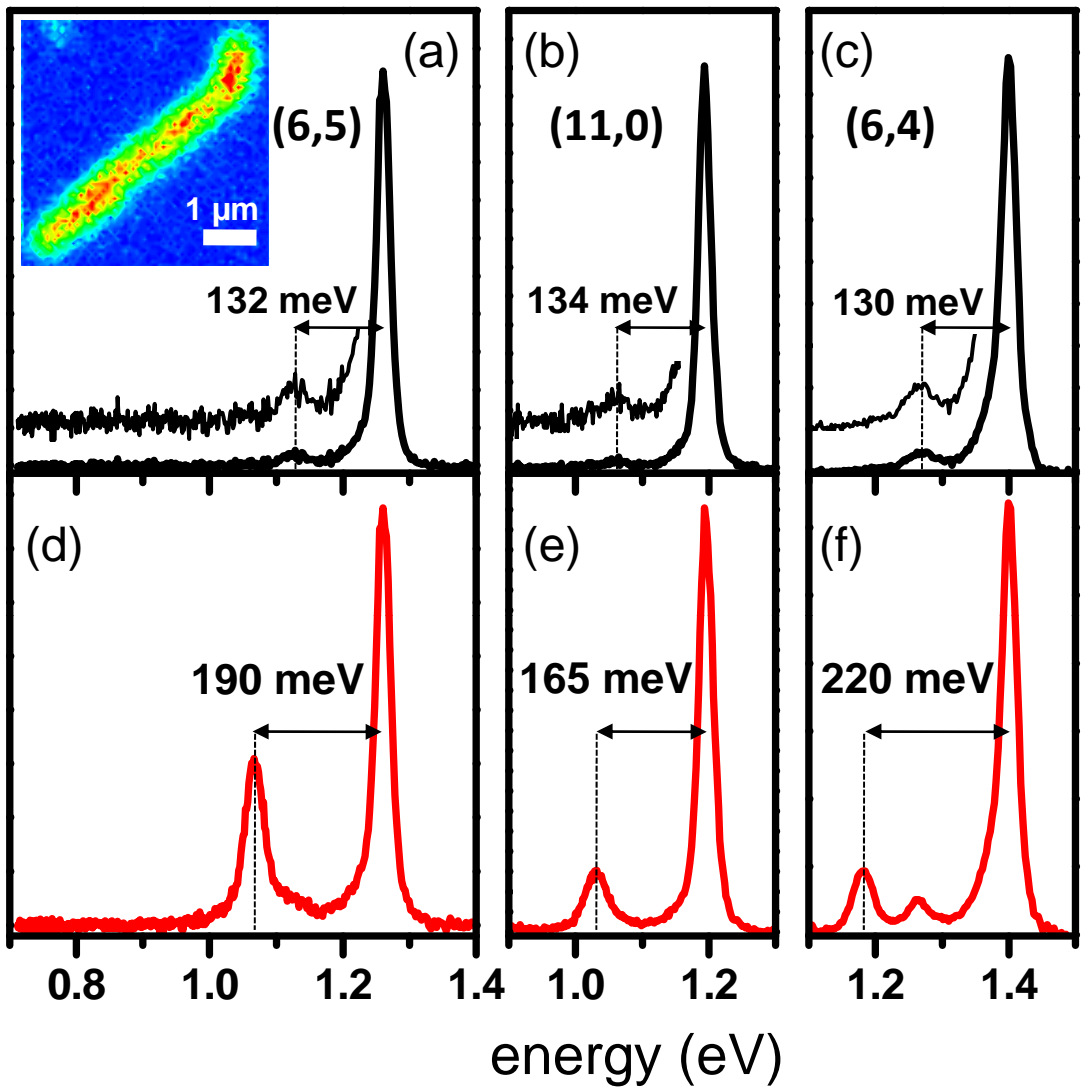

Figure 1

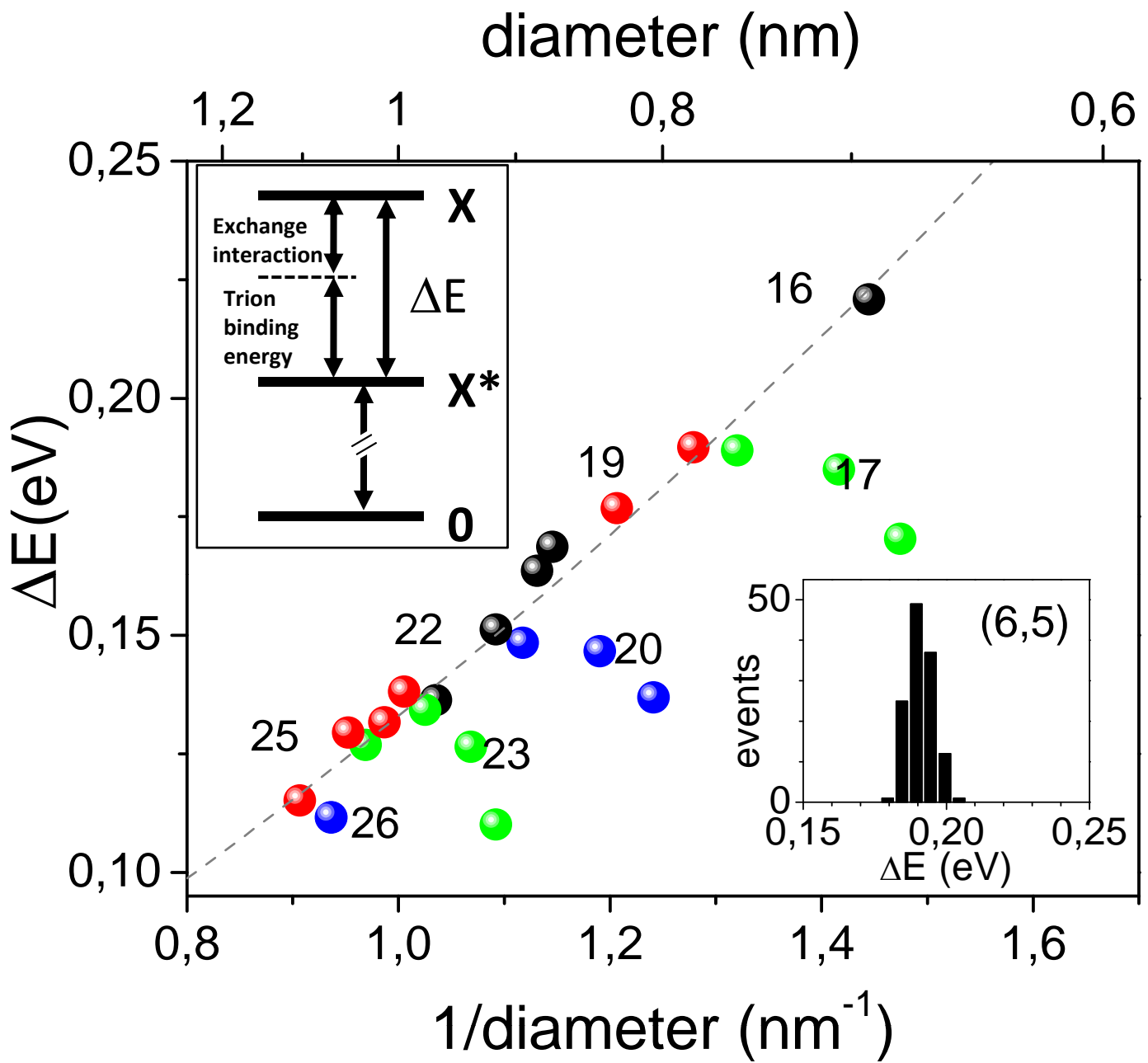

Figure 2

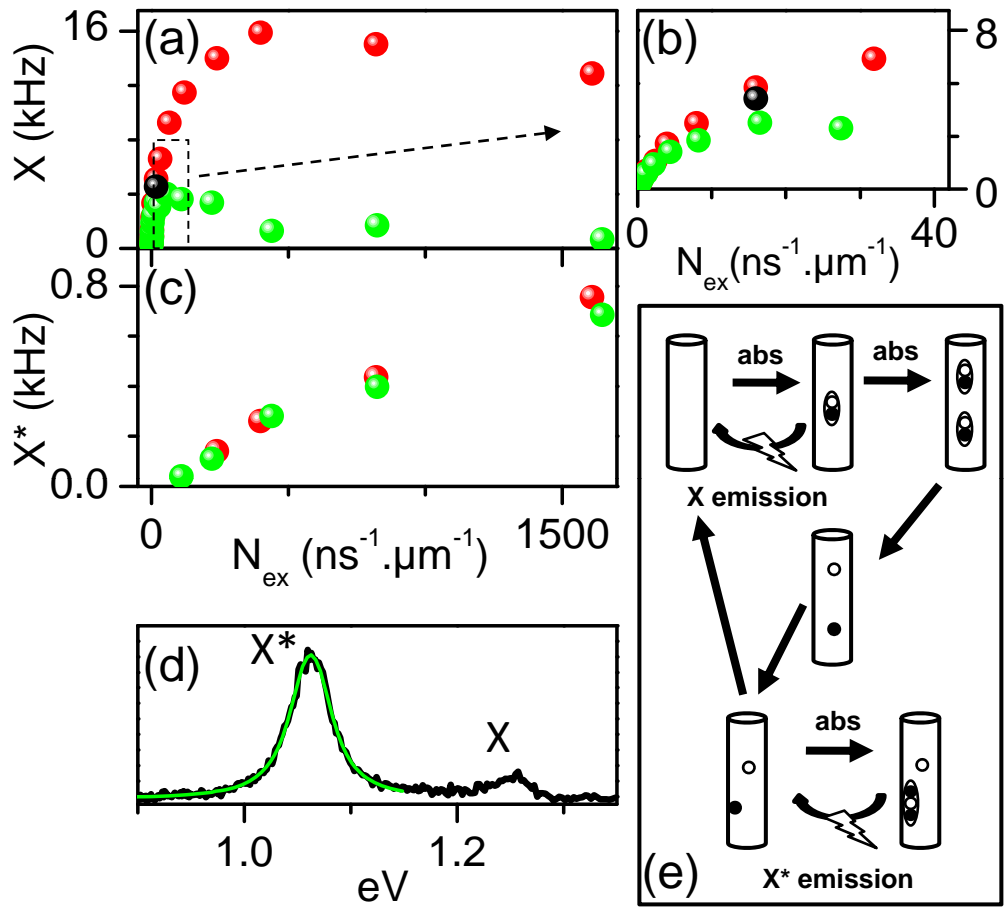

Figure 3

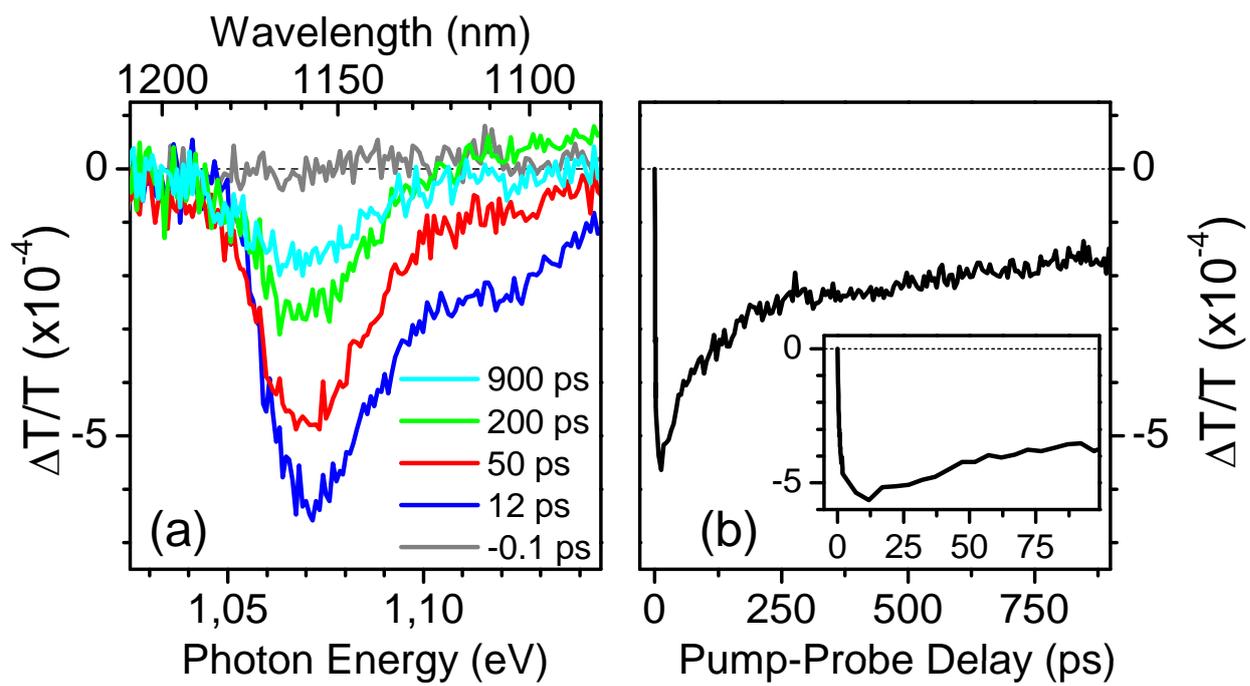

Figure 4